\documentclass[12pt]{article}

\def\4he{\,{{}^4{\rm He}}}
\def\3he{\,{{}^3{\rm He}}}
\def\li7{\,{{}^7{\rm Li}}}

\def\qso{\,{\rm QSO}}
\def\mev{\,{\rm MeV}}
\def\sec{\,{\rm sec}}
\def\nutau{\,{\nu_{\tau}}}
\def\numu{\,{\nu_{\mu}}}
\def\nue{\,{\nu_e}}
\usepackage{latexsym, epsf}

\begin{document}

\newcommand{\gtrsim}{ \mathop{}_{\textstyle \sim}^{\textstyle >} }
\newcommand{\lesssim}{ \mathop{}_{\textstyle \sim}^{\textstyle <} }

\begin{titlepage}

\begin{center}

\vspace{2.5cm}

{\large\textbf{Massive Decaying $\tau$ Neutrino and Big Bang Nucleosynthesis}}

\vspace{1cm}

{\large  M.~Kawasaki$\ ^a$,
K.~Kohri$\ ^{a,}$\footnote{kohri@icrr.u-tokyo.ac.jp}
and Katsuhiko Sato$\ ^b$} \\

\vspace{1cm}

$^a$\textit{Institute for Cosmic Ray Research, University of Tokyo,
  Tanashi, Tokyo 188, Japan} \\

$^b$\textit{Department of Physics and Research Center for the Early
Universe, School of Science, University of Tokyo, Tokyo 113, Japan}

 
\end{center}

\vspace{3cm}

\begin{abstract}
    Comparing Big Bang Nucleosynthesis predictions with the light
    element abundances inferred from observational data, we 
    can obtain the strong constraints on some neutrino properties, e.g.
     number of neutrino species, mass, lifetime. Recently the
     deuterium abundances were measured in high red-shift QSO
     absorption systems. It is expected that they are close to the
    primordial values, however, two groups have  reported inconsistent
    values which are different in one order of magnitude. In this
    paper we show how we can constrain on $\tau$ neutrino mass and 
    its lifetime in each case when we adopt either high or low
    deuterium data. We find that if $0.01 \sec
    \lesssim \tau_{\nutau} \lesssim 1 \sec$ and $10\mev \lesssim
    m_{\nutau}  \lesssim 24\mev$, the theoretical predictions agree
    with the low D/H abundances. On the other hand if we adopt the
    high D/H abundances, we obtain the upper bound of $\tau$ neutrino
    mass, $m_{\nutau}\lesssim 20 \mev$.

\end{abstract}

\end{titlepage}
\newpage

\section{Introduction}

 It is a very important problem for particle physicists and
 astrophysicists whether a neutrino has a finite mass or
 not. Neutrinos are likely massive since it is known that neutrinos
 should have a mass to solve some astrophysical problems such as solar
 neutrino problem (e.g,~\cite{Bahcall}).  Among three species of
 neutrinos, it is presumed that $\tau$ neutrino is heaviest. The
 present experimental upper bound on the mass of $\tau$ neutrino is
 \cite{pdg} 
\begin{equation}
    \label{experiment}
    m_{\nutau} < 24 \mev \ \ ( 95 \% \ {\rm C.L.}).
\end{equation}
 On the other hand, the mass of a stable $\tau$ neutrino should be
 less than about 100eV otherwise its cosmic density exceeds the
 critical density of the universe. Therefore massive $\tau$ neutrino
 with mass between 100eV and 24MeV must be unstable, i.e., decays into
 other particles. However the radiative decay of a massive neutrino is 
 stringently constrained by laboratory experiments, astrophysics and
 cosmology and only non-radiative decay is allowed.  

 Big Bang Nucleosynthesis can strongly constrain on such neutrino
 properties, in particular  the number of neutrino species, mass,
 lifetime, chemical potential and so on,  when we compare the
 theoretical prediction of light elements ($\4he$, D, $\3he$ and
 $\li7$) with the primordial value inferred from the observational
 data. For example, since neutrino is one of the dominant components
 of the energy density in the universe at the BBN epoch, the change of
 neutrino energy density affects the Hubble expansion rate and alters
 the freeze out time of neutron to proton ratio. Then the predicted
 light element abundances, especially $\4he$ in this case, are
 sensitively changed.  More precisely we measure the primordial
 component of light element abundances, more strongly we constrain on
 such neutrino properties. 

 Recently D abundances were measured in high red-shift QSO
 absorption systems which would not suffer from the effects of the
 chemical evolution and would have D abundance which is closer to the
 primordial value. However, two groups have  reported inconsistent D/H
 values which are different in one order of magnitude. Since we do not
 have any definite models of the galactic and  stellar chemical
 evolutions, it may be premature for us to judge which measurements are
 more correct.  

 If the high D abundances, D/H =
 $(1.9\pm0.5)\times10^{-4}$~\cite{songaila}, in high red-shift QSO
 absorption systems are adopted as the  primordial values, since they
 agree with other light element measurements,  we can obtain
 the rigid constraints for some model parameters. On the other hand,
 if the low D abundances,  D/H =
 $(3.39\pm0.25)\times10^{-5}$~\cite{tytler1, tytler2}, are adopted, there is a
 discrepancy between the standard BBN theory and the observational
 measurements. This leads to the similar problem which was pointed out
 by Hata et al. (1995)~\cite{hata} though they used the data for D and
 $^3$He in solar neighborhood and a simple chemical evolution model to
 infer the primordial abundances. They found that theoretical
 predictions agree with the observational data if the number of
 neutrino species $N_{\nu}$ is less than 2.6.  The point is that the
 prediction of the standard ($N_{\nu}=3$) BBN for $^4$He is too large
 if the photon-baryon ratio $\eta$ is chosen to fit the D data ( or
 the prediction of D is too large if $\eta$ is determined by $^4$He
 data). In this  case, we should look for non-standard scenarios.
 
 There are a couple of ideas proposed to solve the discrepancy. One
 can reduce the abundance of $^4$He by assuming non-standard
 properties of neutrinos (e.g. degenerate electron
 neutrino~\cite{kohri}, massive neutrino ) or can decrease the D
 abundance by decays of exotic particles~\cite{HoKaMo}. In particular,
 it was pointed out~\cite{KaStKa,kernan} that a massive and short
 lived $\tau$ neutrino with mass $O$(MeV) and lifetime about $0.1$
 sec. can reduce the effective number of neutrino species and hence
 the abundance of $^4$He. In this paper we study the effect of a
 massive and short lived $\tau$ neutrino on the BBN with a statistical
 analysis. Compared with the recent observational data, especially
 adopting the low D/H in high z QSO absorption systems as a deuterium
 value, we find that if $0.01 \sec  \lesssim \tau_{\nutau} \lesssim
 1\sec$ and $10\mev \lesssim m_{\nutau} \lesssim 24\mev$ the
 theoretical predictions agree with the all  observational 
 data. In the case where we choose the high D abundance, we find 
 the upper bound of the $\tau$ neutrino mass, $m_{\nutau} \lesssim
 20\mev$. 


\section{Cosmological Evolution of Massive Neutrino}

We study the BBN effects of the decaying $\tau$ neutrino by solving
a set of Boltzmann equations which describe the evolution of momentum
distributions of the $\tau$ neutrino and its decay products. A Dirac
$\nu_{\tau}$ with mass larger than few tens of keV is not allowed
since it cools SN1987A too fast. Thus, in this paper we only consider
a Majorana $\tau$ neutrino. We take into account the three
interactions, (1) decay process, (2) pair annihilation and creation
process, and (3) scattering process. Concerning the decay process, we
suppose that a massive Majorana $\tau$ neutrino with mass $m_{\nutau}$
and lifetime $\tau_{\nutau}$, decays into a $\mu$ neutrino and a scalar
particle:\footnote{
The $\tau$ neutrino might decay into a electron neutrino which changes the
neutron-proton ratio directly. However, in the present paper we do not 
consider this decay channel since the Boltzmann equation for $\nu_e$ is 
too complicated. It is expected the emissions of $\nu_e$ and
$\bar{\nu_e}$ increase
the neutron-proton ratio and tend to enhance the discrepancy between
the BBN theory and the observation. 
}
\begin{equation}
    \nu_{\tau} \rightarrow \nu_{\mu} + \phi,
\end{equation}
where we assume that $\phi$ and $\nu_{\mu}$ are massless. The scalar
particle may be a Majoron~\cite{majoron} or a
Familon~\cite{familon} and the decay modes are expected
to be detected in the future experiments, CLEO III or B factories
~\cite{feng}. 
The Boltzmann equation for the distribution $f_i$ of $i$-particle ($i=
\nu_{\tau}, \nu_{\mu}, \phi$) is written
as~\cite{kolbturner,bernstein},
\begin{equation}
    \label{bol}
    \frac{\partial f_i}{\partial t} -
    \frac{\dot{a}(t)}{a(t)}p_i\frac{\partial f_i}{\partial p_i} =
    C_i(decay) + C_i(ann) + C_i(scatt)
\end{equation}
where $a(t)$ is the scale factor, $p_i$ is the momentum, and $C_i$s are
collision terms which correspond to the three processes. 

The collision term $C_{\nu_{\tau}}$ for the decay process
$\nutau(p_{\nutau}) \to \numu(p_{\numu}) + \phi(p_{\phi})$ is given
by~\cite{KaStKa}
\begin{equation}
    \label{decay}
    C_{\nutau}(decay) = - \Gamma_D^{\nutau} + \Gamma_{ID}^{\nutau}
\end{equation}
where $\Gamma_D^{\nutau}$ and $\Gamma_{ID}^{\nutau}$ are  the 
collision terms for the decay and the inverse decay processes,
respectively. They
are given by,
\begin{eqnarray}
    \label{taudec1}
    \lefteqn{\Gamma_D^{\nutau}}\\ &=&\nonumber
    \frac{1}{\tau_{\nutau}}\frac{m_{\nutau}}{\sqrt{m_{\nutau}^2 +
    p_{\nutau}^2}p_{\nutau}}f_{\nutau}(p_{\nutau})\int_{p_{mim}}^{p_{max}}
    dp_{\phi} \left(1 +
    f_{\phi}(p_{\phi})\right)\left(1 - f_{\numu}(\sqrt{m_{\nutau}^2 +
    p_{\nutau}^2} - p_{\phi})\right),\\
    \label{taudec2}
    \lefteqn{\Gamma_{ID}^{\nutau}}\\ &=&\nonumber
    \frac{1}{\tau_{\nutau}}\frac{m_{\nutau}}{\sqrt{m_{\nutau}^2 +
    p_{\nutau}^2}p_{\nutau}}(1 - f_{\nutau}(p_{\nutau}))
    \int_{p_{mim}}^{p_{max}}dp_{\phi} f_{\phi}(p_{\phi})
    f_{\numu}(\sqrt{m_{\nutau}^2 +p_{\nutau}^2} - p_{\phi}).
\end{eqnarray}
The integration is performed between $p_{max} = {(\sqrt{m_{\nutau}^2 +
p_{\nutau}^2} + p_{\nutau})/2}$ and $p_{mim} = (\sqrt{m_{\nutau}^2 +
p_{\nutau}^2} - p_{\nutau})/2$. In the same way, the collision terms
$C_{\nu_{\mu}}(decay)$ and $C_{\phi}(decay)$ are given by,
\begin{equation}
     \label{col}
    C_{\nu_{\mu},\phi}(decay) = \Gamma_D^{\nu_{\mu},\phi}
    - \Gamma_{ID}^{\nu_{\mu},\phi},
\end{equation}
where $\Gamma_D^{\nu_{\mu},\phi},\Gamma_{ID}^{\nu_{\mu},\phi}$ are
given by 
\begin{eqnarray}
    \label{dec1}
    \Gamma_{D}^k &=&\frac{g_{\nutau}}{g_k}
    \frac{1}{\tau_{\nutau}}\frac{m_{\nutau}}{p_k^2}(1 \pm f_k(p_k))
    \int_{m_{\nutau}^2/4p_k}^{\infty}dp_{k'} 
    f_{\nutau}(\sqrt{(p_{\phi}+p_{\numu})^2 - m_{\nutau}^2})
    (1 \pm f_{k'}(p_{k'})),\\
    \label{dec2}
    \Gamma_{ID}^k &=&\frac{g_{\nutau}}{g_k}
    \frac{1}{\tau_{\nutau}}\frac{m_{\nutau}}{p_k^2}f_k(p_k)
    \int_{m_{\nutau}^2/4p_k}^{\infty}dp_{k'}( 1 \pm 
    f_{\nutau}(\sqrt{(p_{\phi}+p_{\numu})^2 -
    m_{\nutau}^2}))f_{k'}(p_{k'}),
\end{eqnarray}
where $ k \ne k'= \numu$ or $\phi$, the spin factors are
$g_{\nutau} = g_{\numu} = 2$, and $g_{\phi} = 1$.
 
For pair annihilation and creation processes $
\nu(p_{\nu}) + \nu(p'_{\nu}) \to l_i(p_i) + l_i(p'_i)$, ($i
= e, \nue, \numu$), the collision terms are written as
\begin{equation}
    \label{ann}
    C(ann) =   - \frac{1}{2\pi^2}\int p'^2_kdp'_k
    (\sigma v)_k (f_k(p_k) f_k(p'_k)- f_{eq}(p_k)f_{eq}(p'_k)),
\end{equation}
where k = $\nutau$, $\numu$, and $f_{eq}$ is the equilibrium
distribution and $(\sigma v)_k$ is the differential cross sections
given by
\begin{equation}
    \label{difan1}
    (\sigma v)_{\nutau} = \frac{G^2_F}{9\pi}\sum_{i = e,\nue,\numu}
    (C^2_{V i} + C^2_{A i})\left[4EE' - \left(\frac{E}{E'} +
            \frac{E'}{E}\right) m^2_{\nutau} - 2
            \frac{m^4_{\nutau}}{EE'} \right],
\end{equation}
\begin{equation}
    \label{difan2}
    (\sigma v)_{\numu} = \frac{4G^2_F}{9\pi}\sum_{i = e,\nue,\numu}
    (C^2_{V i} + C^2_{A i})EE',
\end{equation}
where $G_F$ is Fermi coupling constant, and $C_A = 1/2, C_V = 1/2 -
2 sin^2\theta_W$ ($\theta_W$: Weinberg angle) for e$^{\pm}$ and $C_A =
1/2, C_V = 1/2$ for neutrinos. We do not include the annihilation and
creation process for the scalar particle $\phi$  assuming  that their
interaction is very week.

As for scattering processes, it is not easy to solve the Boltzmann 
equations including the collision terms due to scattering because 
they are generally much more complicated than collision terms due to the
annihilation and the decay.  Without scattering, the spectrum of the 
$\tau$ neutrino is distorted because of the momentum dependence of the 
annihilation rate, which leads to overestimation of the density of 
$\nu_{\tau}$~\cite{KaStKa}.  This effect is significant for the 
neutrino whose mass is larger than about 10MeV since such heavy 
neutrino becomes non-relativistic before decoupling and its density 
decreases through annihilation processes.  On the other hand, for 
neutrinos which are relativistic at decoupling epoch, the momentum 
distributions do not change by the annihilation and is the same as the 
equilibrium one.  Thus the scattering process is important only for 
the heavy $\tau$ neutrino ($m_{\nu_{\tau}}\gtrsim 10$MeV).  Therefore we 
consider only the scattering process $ \nutau(p) + l(k) \to 
\nutau(p') + l(k')$ in the collision term of the Boltzmann equation 
for $\nu_{\tau}$. Using non-relativistic approximation, the collision 
term is simplified as
\begin{equation}
    \label{scatt}
    C(scatt) = -\Gamma_f + \Gamma_b,
\end{equation}
where $\Gamma_f$ and $\Gamma_b$ are forward and backward reaction
rates which are given by
\begin{eqnarray}
    \label{fscatt}
    \Gamma_f = \sum_{i = e,\nue,\numu}\frac{G^2_F}{\pi^3}(C^2_{V i}
                 + C^2_{A
                 i})m^5_{\nutau}\frac{1}{x^3}\frac{e^{\sqrt{x^2 +
                 y^2}/2}}{y\sqrt{x^2 + y^2}}f_{\nutau}(y) \\
                 \nonumber \times\int
                 ^{\infty}_0 \frac{y'dy'}{\sqrt{x^2 +
                 y'^2}}e^{-\sqrt{x^2 + y'^2}/2}r(y,y')(1-f_i (y')),
\end{eqnarray}
\begin{eqnarray}
    \label{bscatt}
    \Gamma_b = \sum_{i = e,\nue,\numu}\frac{G^2_F}{\pi^3}(C^2_{V i}
                 + C^2_{A
                 i})m^5_{\nutau}\frac{1}{x^3}\frac{e^{- \sqrt{x^2 +
                 y^2}/2}}{y\sqrt{x^2 + y^2}}(1 - f_{\nutau}(y)) \\
                 \nonumber \times\int
                 ^{\infty}_0 \frac{y'dy'}{\sqrt{x^2 +
                 y'^2}}e^{\sqrt{x^2 + y'^2}/2}r(y,y')f_i (y'),
\end{eqnarray}
where $x = m_{\nutau}/T, \  y = p/T$, and $y' = p'/T$. The function $r(y,y')$
is given by
\begin{eqnarray}
    \label{rscatt}
    r(y,y') = \left(4 + \frac32 |y - y'| + \frac14 |y - y'|^2\right)
    e^{-\frac{|y-y'|}{2}} \\ \nonumber
    - \left(4 + \frac32 (y + y') + 
    \frac14 (y + y')^2\right) e^{-\frac{y + y'}{2}}.
\end{eqnarray}

At T $\lesssim$ 0.5 MeV, electrons and positrons begin to annihilate
and heat photons without heating neutrinos. Then we have to deal with
the temperature and cosmic time relation carefully. In this
situation, we solve a set of Boltzmann equations coupled with the Hubble
expansion rate and the entropy conservation equation.  The Hubble
expansion rate is obtained by solving the Einstein equation:
\begin{equation}
    \label{hubble}
    \frac{\dot{a}(t)}{a(t)} = \left[\frac{8\pi G}{3}
     (\rho_e + \rho_{\gamma}+ \rho_{\nue}+ \rho_{\numu}+ \rho_{\nutau}+ 
     \rho_{\phi})\right]^{\frac12},
\end{equation}
and the entropy conservation is:
\begin{equation}
    \label{entropy}
    \frac{11}{3} T^3_{\nu}= \frac{4}{3}T^3_{\gamma}\left[1 +
        \frac34\left(\frac{\rho_e + P_e}{\rho_{\gamma}}\right)\right],
\end{equation}
where $\rho_i$ denotes the energy density of the particle ``$i$'' ($i=
e, \gamma, \nu_{e}, \nu_{\mu}, \nu_{\tau}, \phi$) and $P_e$ is the
electron pressure.

As initial conditions we take
\begin{eqnarray}
    f_{\nu_{\tau}} = f_{\nu_{\mu}} = f_{eq}\\
    f_{\phi} = 0,
\end{eqnarray}
where we assume that the scalar particle has been never thermalized and
its cosmological density is negligible. We will discuss the justification
of this assumption later.

\section{Time Evolution of $N_{\nu}$}

As we have already mentioned, the abundances of $\4he$ is sensitive to
the total energy density of the universe during the BBN epoch, 
T = $10\mev -
0.1\mev$. The lower the energy density, the more slowly the universe
expands, which leads to later decoupling of the weak interactions and
less neutron-proton ratio. Then the abundance of $\4he$ decreases
since almost all neutrons are translated into $\4he$. Since the other
light elements are insensitive to the cosmic expansion rate, it is
expected that the BBN predictions may agree with observational data if
the energy density of the universe becomes lower than the standard one
by the effect of heavy $\tau$ neutrino decay.
 
When we discuss the energy density of massive decaying $\tau$ neutrino,
it is useful to introduce the effective number of neutrino species
$N_{\nu}$ defined by
\begin{equation}
    N_{\nu} = \frac{\rho_{\nue} + \rho_{\numu} + \rho_{\nutau} +
    \rho_{\phi}}{\rho_{\nu}(m_{\nu}=0)}.
\end{equation}
The denominator is the energy density of a massless neutrino and
$N_{\nu}$ is a constant ( = 3 ) in the standard model.

In Fig.~1 we show the time evolutions of $N_{\nu}$ for
$(\tau_{\nu_{\tau}}(\sec), m_{\nu_{\tau}}(\textrm{MeV})) = (10^{-3}, 1),
(1, 10)$ and $(0.1, 14)$. In order to understand their behaviors, it
is convenient to 
introduce the three temperature scales, (1) the epoch when the
decaying particle becomes non-relativistic $T_{NR}$, (2) the decaying
particle's decoupling epoch $T_D$, and (3) the epoch when the decay
occurs $T_{decay}$. We define them by,
\begin{eqnarray}
     \label{timescale}
     T_{NR} &\equiv& \frac{m_{\nutau}}{3},\\
     \Gamma(T_D) &\equiv& H(T_D),\\
     T_{decay} &\equiv& T(t = \tau_{eff}),
\end{eqnarray}
where $H(t)$ is the Hubble expansion rate, and $\tau_{eff}$ is the
effective lifetime of the $\tau$ neutrino. Notice that the effective
lifetime of the relativistic neutrino becomes longer than the proper
lifetime because of the time dilation ( $\tau_{eff} \sim \tau
\sqrt{m_{\nutau}^2 + p_{\nutau}^2}/m_{\nutau}$). 

In the case of $\tau_{\nutau} = 10^{-3}$ sec and $m_{\nutau} = 1$ MeV,
the relation of the three time scales is given by $T _{decay} >T_{D}
>T_{NR}$. At temperatures between $T_{decay}$ and $T_{D}$ the decay,
inverse decay, annihilation and pair creation processes tend to
establish the thermal equilibrium among three species of neutrinos and
the scalar particle and $N_{\nu}$ becomes close to 3.6. Then, after
decoupling, $N_{\nu}$ increases due to the mass effect until $T_{NR}$
at which the inverse decay becomes suppressed and $\nu_{\tau}$ begins
to decay efficiently. On the other hand, in the case of $\tau_{\nutau}
= 1$ sec and $m_{\nutau} = 10 \mev$, $T_{NR} >T_{D} >
T_{decay}$. Between $T_{NR}$ and $T_{D}$ the energy density of $\tau$
neutrino decrease due to the Boltzmann suppression. After decoupling,
the mass effect increases $N_{\nu}$, and at $T\sim T_{decay}$
$\nu_{\tau}$ begins to decay.  Thus, if $T_{D} \sim T_{decay}$, the
increases of $N_{\nu}$ due to the mass effect is not significant and
N$_{\nu}$ becomes less than three as shown in Fig.~1 for
$\tau_{\nu_{\tau}}=10^{-1}$ sec and $m_{\nu_{\tau}}= 14$ MeV.

\section{BBN with Massive Decaying $\tau$ Neutrino}

In a previous section, we have discussed the time evolution of the
energy density in the universe, and have shown that $\tau$ neutrino
with $O(10\mev)$ mass and short lifetime ($\sim 0.1$ sec) can reduce
the effective number of neutrino species $N_{\nu}$. In order to
calculate the abundances of D, $\4he, \3he$, and $\li7$, we
incorporate the change of the effective number of neutrino species
N$_{\nu}$ into the big bang nucleosynthesis code~\cite{kernan}. 

In the case of massive and unstable tau neutrino, the effective number
of neutrino species is a time-dependent variable. Since only
$\4he$ abundance is sensitive to N$_{\nu}$, if the
predicted $\4he$ abundance is expressed by the corresponding (constant)
number of neutrino species in the BBN calculation without massive
neutrinos, it helps us to understand the results of our calculation
intuitively.
Thus, we introduce the equivalent number of neutrino species 
$\bar{N_{\nu}}$ here which is defined as
\begin{equation}
    \label{equiv}
    Y_p( \bar{N_{\nu}}, m_{\nutau}=0, \tau_{\nu_{\tau}}=\infty) \equiv
    Y_p(N_{\nu}=3, m_{\nutau}, \tau_{\nu_{\tau}}).
\end{equation}
 In Fig.~2 we show the contour plots of the equivalent number of
 neutrino species $\bar{N_{\nu}}$ in the parameter space,
 $\tau_{\nutau}-m_{\nutau}$. In this plot  we expect that the predicted
 Y$_p$ decreases if $\tau_{\nutau} \lesssim 10\sec$ and
 $m_{\nutau} \gtrsim 10 \mev$.

In the
numerical calculation we make use of the Monte Carlo procedure in
order to include the experimental uncertainties of the nuclear
reaction rates and the neutron lifetime~\cite{kawano,smith}. Since the
theoretical $1\sigma$ errors are almost independent of $N_{\nu}$ for
$N_{\nu} = 2- 4$, we adopt the $1\sigma$ errors estimated by the
standard BBN ($N_{\nu}=3$) calculation. 

The calculated abundances are compared with the primordial values
inferred form the observational data. Concerning the $\4he$, we adopt
the value in the extra galactic HII region and it is given
by~\cite{olive},
\begin{equation}
    Y_p = 0.234  \pm  0.002(\mbox{stat}) \pm 0.005(\mbox{syst})
\end{equation}
where $Y_P$ is the mass fraction of $\4he$ and ``p'' denotes the
primordial value.  It is believed that abundances of $^7$Li observed
in the population II metal poor halo stars in our galaxy represent the
primordial values. We adopt the recent observational data~\cite{boni},
\begin{equation}
 (\li7/{\rm H})_p = (1.73 \pm 0.05(\mbox{stat}) \pm 0.20(\mbox{syst}))
                 \times 10^{-10}.
\end{equation}

For a deuterium abundance, we have some types of the
measurements which  are the local ISM or the solar system
data and the abundances observed in high red-shift QSO absorption
systems. The abundances inferred from the local ISM are
$({\rm D/H})_{ISM} = (1.6 \pm 0.2) \times 10^{-5}$~\cite{linsky}.
In the solar system the observed deuterium abundances are
$({\rm D/H})_{\odot} = (2.57 \pm 0.92) \times 10^{-5}$~\cite{geiss}. 
This pre-solar deuterium abundance is obtained  by subtracting  $\3he$/H value
observed in carbonaceous chondorites from (D +$\3he$)/H  observed 
in the solar winds. From above  abundances, we must
extrapolate the true primordial deuterium abundance assuming 
appropriate chemical evolution models. 

On the other hand, The observations in high red-shift QSO absorption
systems are expected 
that they are nearly the primordial values  because such Lyman
$\alpha$  clouds would be unprocessed. In the
last three years, more papers concerning these Lyman limit system
observations are published. However, two groups have measured
different D/H values which disagree in a order of magnitude.
The higher values measured by some persons  are~\cite{songaila},
\begin{equation}
    \label{songaila}
     ({\rm D/H})_{\qso} =  (1.9 \pm 0.5) \times 10 ^{-4}.
\end{equation}
The lower values measured by Burles and Tytler.~\cite{tytler2} are,
\begin{equation}
    \label{tytler2}
    ({\rm D/H})_{\qso} =  (3.39 \pm 0.25) \times 10 ^{-5}.
\end{equation}

Some chemical evolution models which predict only
modest destruction of deuterium during the galaxy
evolution  have been suggested ~\cite{tosi, chiappini}. Adopting such
a recent model 
~\cite{chiappini} 
which reports that the deuterium destruction is limited to a factor 3
or less,  the above low deuterium
values in high red-shift QSO absorption systems would be consistent with
the ISM and solar system data. However, there are also the other
models in which primordial deuterium can be significantly destroyed
into a order of magnitude lower without over-producing the $\3he$
~\cite{scully}.  The latter models might agree with the high deuterium 
abundance in high z QSO absorption systems. 
Considering the circumstances mentioned above, we so far have no
conclusive model of the galactic chemical evolution. On the
observational side, it is premature to determine which
deuterium values in high red-shift QSO absorption systems are
reliable. Therefore in this paper, we adopt both  the high and low
deuterium data in high z QSO absorption systems and for each deuterium 
value we constrain on the $\tau$ neutrino mass and its lifetime.


In order to obtain the confidence level of the model parameters, mass
$m_{\nutau}$, lifetime $\tau_{\nutau}$, and baryon to photon ratio
$\eta$ in the universe, we utilize the Maximum Likelihood analysis. The
likelihood function of theoretical abundances is taken to be gaussian
which has the one sigma variance calculated by the Monte Carlo
procedure. The likelihood function of observational data is taken to
be gaussian for the statistical error, and to be top hat distribution
for the systematic error (see Ref.~\cite{kohri}). Also for the $\tau$
neutrino mass, the experimental bound Eq.~(\ref{experiment}) is
incorporated as a gaussian distribution.

 The obtained two dimensional contours in the parameter space,
 $\tau_{\nutau}-m_{\nutau}$, are shown in Fig.~3 for using the low 
 (D/H)$_{\qso}$, and in Fig.~4 for using the high
 (D/H)$_{\qso}$. In Fig.~3 the contour with same likelihood as the
 standard BBN model (N$_{\nu}$ = 3) is shown by the dotted
 line. Notice that if  
 $0.01\sec  \lesssim \tau_{\nutau} \lesssim 1\sec$ and $10\mev \lesssim
 m_{\nutau} \lesssim 24\mev$, the theoretical predictions give a much
 better fit to the observational data than the standard BBN.  The best
 fit values are estimated as 
\begin{eqnarray}
    \label{qtaumas}
    m_{\nutau} &=& 15^{\ +11}_{\ -5} \ \mev 
                           \qquad\quad (68\% \ {\rm C.L.}),\\
    \label{qtau}
    \log_{10}\left(\frac{\tau_{\nutau}}{{\rm sec}}\right) &=& 
       - 1.0^{\ +1.0} _{\ -1.2} \ \qquad\qquad (68\% \ {\rm C.L.}),
\end{eqnarray}
for the low D abundances in the QSO absorption systems. The filled
square in Fig.~3 represents the best fit point.
From  Fig.~4 we can obtain the upper limit for the $\tau$
neutrino mass $ m_{\nutau}$,
\begin{equation}
    \label{upper}
    m_{\nutau} \lesssim 20 \mev \qquad\quad (68\% \ {\rm C.L.}),
\end{equation}
for high D abundances in the QSO absorption systems.

\section{Conclusion and  Discussion}

In this paper we have shown the light element abundances which are
predicted by BBN theory with $\tau$ neutrino whose  mass is $\sim 10$MeV
and lifetime is $\sim 0.1$ sec. agree with the primordial values inferred
by observational data even if we adopt the low D/H in high z QSO absorption 
systems. On the other hand, if we adopt the high D/H, we can obtain the 
strong constraints on $\tau$ neutrino mass $m_{\nutau}\lesssim 20$MeV.

Finally we discuss the annihilation rate for the scalar particle
$\phi$. For this purpose we assume that $\phi$ is a Majoron-like
particle which has the interaction  to $\nu_{\tau}$ as
\begin{equation}
    L_{\rm int} = \frac{1}{v}\partial_{\mu}\phi
    \bar{\nu}_{\tau}\gamma_5\gamma^{\mu}\nu_{\tau},
\end{equation}
where 
$v$ is the characteristic
energy scale. In the relativistic limit the ratio of the annihilation
rate for $\phi + \phi \rightarrow \nu_{\tau}+\nu_{\tau}$ to Hubble expansion
rate is given by
\begin{equation}
    \label{tot_crs}
    \frac{\Gamma}{H} = \sigma v_{rel} n_{\phi} /H \simeq
1 \times 10^{20}\left(\frac{m_{\nutau}}{v}\right)^4
\left(\frac{T}{\mev}\right)^{-1} 
\log\left(\frac{1.21 T}{m_{\nutau}}\right) \quad (T \gg m_{\nutau}).
\end{equation}
When the cosmic temperature $T$ is less than $m_{\nutau}$, the ratio 
$\Gamma / H \propto T^2$. Thus $\Gamma / H$  takes the
maximum value at $T \simeq m_{\nutau}$:
\begin{equation}
    \label{max}
    \left(\frac{\Gamma}{H}\right)_{\rm max} \simeq 2 \times 10^{19}
    \left(\frac{\mev}{v}\right)^4.
\end{equation}
The condition that the scalar particle
$\phi$ is not in thermal equilibrium is given by
\begin{equation}
    \label{phi-anni}
    v \gtrsim 7 \times 10^4 \mev
\end{equation}
On the other hand, the lifetime of the $\tau$ neutrino is related to
$ v $ by
\begin{equation}
    \tau_{\nu_{\tau}} \simeq 
    \frac{16\pi v^2}{m_{\nutau}^3|U_{\tau,\mu}|^2},
\end{equation}
where $U_{\tau,\mu}$ is the mixing angle. 
Therefore the condition (\ref{phi-anni}) is written as 
\begin{equation}
    \label{lifecrt}
    \tau_{\nu_{\tau}} \gtrsim \ 4 \times 10^{-10} {\rm sec} 
 \ |U_{\tau,\mu}|^{-2}.
\end{equation}
From Eq.~(\ref{lifecrt}) the scalar particles are not in thermal
 equilibrium if $\tau_{\nu_{\tau}} \gtrsim \ 1 \times 10^{-5}$sec
 and $|U_{\tau,\mu}| \gtrsim 10^{-2}$.
 Thus our assumption
that the scalar particle has not been thermalized is justified when
the mixing angle is large. 

 We wish to thank T. Asaka for useful discussions for the Majorana
 neutrino. This work has been supported by the Grant-in-Aid for COE
 Research (07CEQ2002) of the Ministry of Education, Science, and
 Culture in Japan. 


\newpage
\noindent{{\large\textbf{Figure caption}}}\\


\vspace{0.5cm}
\noindent{Figure1: Evolution of the effective number of neutrino
species $N_{\nu}$
    with the cosmic temperature T for $(\tau_{\nu_{\tau}}(\sec), m_{\nu_{\tau}}(\textrm{MeV})) = (10^{-3}, 1),
(1, 10)$ and $(0.1, 14)$.}\\

\vspace{1cm}
\noindent{Figure2: Contours of the equivalent number of neutrino 
    species $\bar{N_{\nu}}$ in the parameter 
    space, $\tau_{\nutau}-m_{\nutau}$. $\bar{N_{\nu}}$ is defined by
    Eq.~(\ref{equiv}). In this plot  we expect that the
    predicted Y$_p$ decreases  if $\tau_{\nutau} \lesssim 10\sec$
    and $m_{\nutau} \gtrsim 10 \mev$. By the definition, 
    $\bar{N_{\nu}}$ does not depend on the baryon to photon ratio
    $\eta$} at all.\\ 

\vspace{1cm}
\noindent{Figure3: Contours of the confidence levels in the parameter
    space, $\tau_{\nutau}-m_{\nutau}$. In this plot the lower 
    deuterium abundance obtained in high red-shift QSO absorption
    systems is adopted as the primordial value. The dashed line
    denotes 90$\%$ C.L., and the solid  line denotes 68$\%$ C.L.. 
     The contour with same likelihood as standard BBN model (N$_{\nu}$
     = 3)  is shown by the dotted line. The filled 
square represents the best fit point. }\\

\vspace{1cm}
\noindent{Figure4: Contours of the confidence levels in the parameter
    space, $\tau_{\nutau}-m_{\nutau}$. In this plot the higher 
    deuterium abundance obtained in high red-shift QSO absorption
    systems is adopted as the primordial value. The dashed line
    denotes 90$\%$ C.L., and the solid  line denotes 68$\%$ C.L.. 
     The contour with same likelihood as standard BBN model (N$_{\nu}$
     = 3)  is shown by the dotted line.}

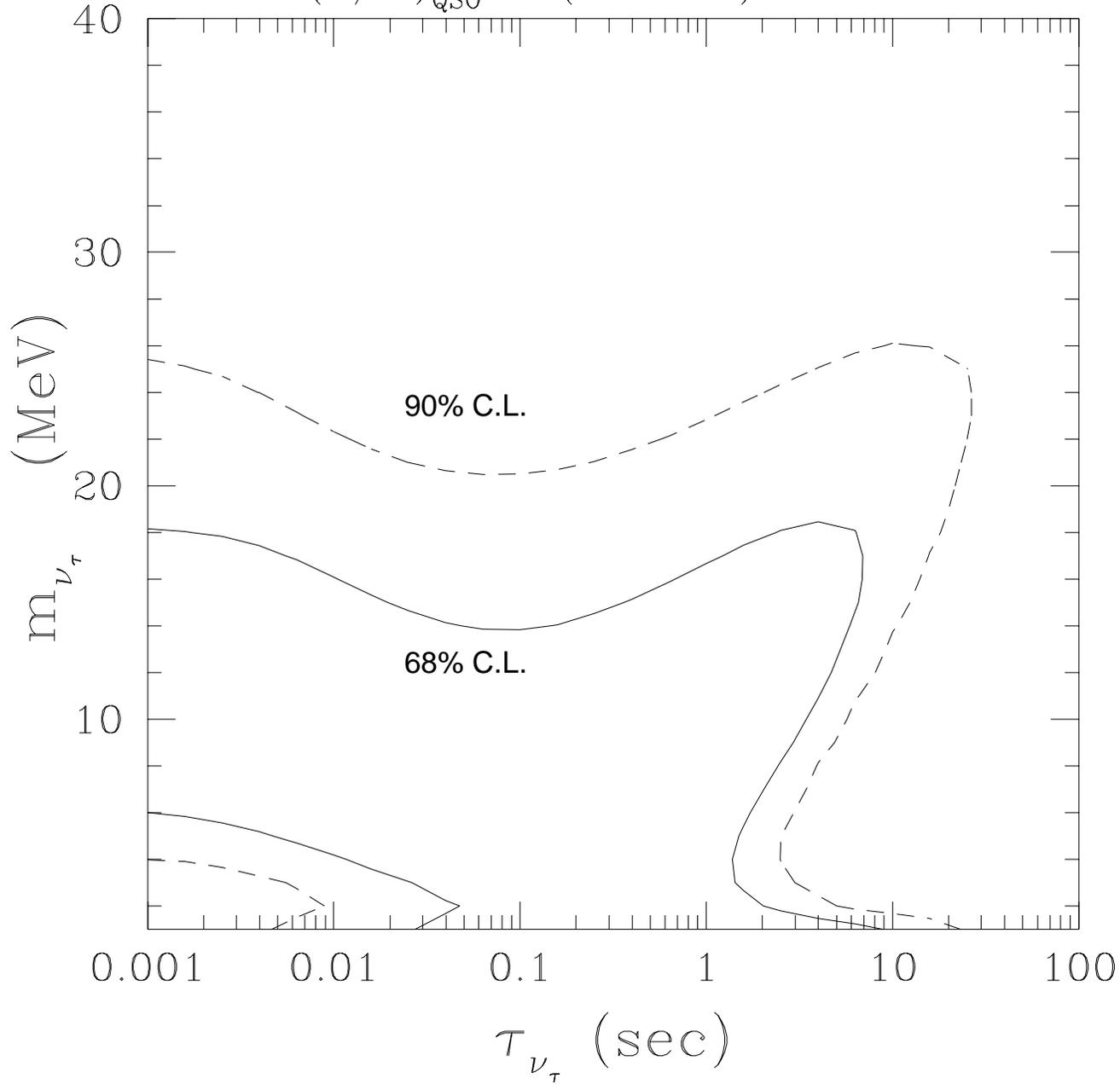
\end{document}